\documentclass[twocolumn, mathline]{aastex631}
\usepackage{hyperref}
\usepackage{amsmath}
\usepackage{float}
\usepackage{placeins}

\newcommand{\Mbmx}{$M_{b} = 4.0 \pm 0.6 ~M_{\oplus}$}
\newcommand{\Mcmx}{$M_{c} = 2.0 \pm 0.7~M_{\oplus}$}
\newcommand{\Mbf}{$M_{b} = 4.1 \pm 0.4~M_{\oplus}$}
\newcommand{\Mcf}{$M_{c} = 2.4 \pm 0.5~M_{\oplus}$}
\newcommand{\rhob}{$\rho_{b}$ = 1.3 $\pm$ 0.4~g~cm$^{-3}$}
\newcommand{\rhoc}{$\rho_{c}$ = 1.4 $\pm$ 0.3~g~cm$^{-3}$}

\begin{document}
\title{Revised Masses for Low Density Planets Orbiting the Disordered M-dwarf System TOI-1266}

\author[0000-0003-0298-4667]{Dakotah Tyler}
\affiliation{Department of Physics and Astronomy, University of California,	Los Angeles, CA 90095, USA}
\author[0000-0003-0967-2893]{Erik A. Petigura}
\affiliation{Department of Physics and Astronomy, University of California,	Los Angeles, CA 90095, USA}
\author[0000-0001-7615-6798]{James Rogers}
\affiliation{Department of Physics and Astronomy, University of California,	Los Angeles, CA 90095, USA}
\author[0000-0001-8342-7736]{Jack Lubin}
\affiliation{Department of Physics and Astronomy, University of California,	Los Angeles, CA 90095, USA}
\author[0000-0003-4526-3747]{Andreas Seifhart}
\affiliation{Gemini Observatory, Hilo, HI 96720, USA}
\author[0000-0003-4733-6532]{Jacob L. Bean}
\affiliation{Department of Astronomy \& Astrophysics, University of Chicago, Chicago, IL, USA}
\author[0000-0003-2404-2427]{Madison Brady}
\affiliation{Department of Astronomy \& Astrophysics, University of Chicago, Chicago, IL, USA}
\author[0000-0002-4671-2957]{Rafael Luque}
\affiliation{Department of Astronomy \& Astrophysics, University of Chicago, Chicago, IL, USA}
\affiliation{NHFP Sagan Fellow}


\begin{abstract}
We present an analysis of 126 new radial velocity measurements from the MAROON-X spectrograph to investigate the TOI-1266 system, which hosts two transiting sub-Neptunes at 10.8 and 18.8 days. We measure masses of $M_{b}=4.01~\pm~0.55~M_{\oplus}$ for TOI-1266 b and  $M_{c}=2.00~\pm~0.72~M_{\oplus}$ for TOI-1266 c. Our mass measurements agree with existing HARPS-N observations which we combined using a weighted average yielding masses for TOI-1266 b, and c of $M_{b}=4.10~\pm~0.43~M_{\oplus}$, $M_{c}=2.4~\pm~0.54~M_{\oplus}$ respectively. The combined dataset enabled a $\approx30\%$ improvement in mass precision. With bulk densities of $\rho_{b}$ = 1.25 $\pm 0.36$~g~cm$^{-3}$ and $\rho_{c}$ = 1.36 $\pm$ 0.31~g~cm$^{-3}$, the planets are among the lowest density sub-Neptunes orbiting an M dwarf. They are both consistent with rocky cores surrounded by hydrogen helium envelopes. TOI-1266 c may also be consistent with a water-rich composition, but we disfavor that interpretation from an Occam's razor perspective.
\end{abstract}


\section{Introduction}
\label{sec:Intro}
One of the key findings from the Kepler mission was that the majority of Sun-like stars host at least one planet between the size of Earth and Neptune with an orbit within 100~days (see Bryson et al. \citeyear{2021AJ....161...36B} and references therein). Notably, such planets are absent from our own solar system. Furthermore, these small close-in planets are shepherded into two distinct radius populations: super-Earths ($<$1.5~$R_{\oplus}$) and sub-Neptunes ($2.0~R_{\oplus}<R_{p}<4.0~R_{\oplus}$). The observed dip in occurrence for planets 1.5--2.0~$R_{\oplus}$ is known as the `radius gap' (Fulton et al. \citeyear{2017AJ....154..109F}; Van Eylen et al. \citeyear{2018MNRAS.479.4786V}) which could be explained by several different potential mechanisms. 

This gap can be explained by a more uniform initial distribution of exoplanets 1--20~$M_{\oplus}$ that undergoes atmospheric mass loss driven by photoevaporation (Lopez \& Fortney \citeyear{2013ApJ...776....2L}; Owen \& Wu \citeyear{2013ApJ...775..105O}, \citeyear{2017ApJ...847...29O}). In this model, X-ray and extreme ultraviolet (XUV) radiation from the host star excites and strips the primordial atmosphere from the planet. While this effect is the strongest early on in a system ($<$ 300~Myr), active photoevaporative mass loss has been detected in mature systems (see Tyler et al. \citeyear{2024ApJ...960..123T} and references therein). 

Another theory explains the radius gap as a consequence of the luminosity of a planet's core. In this `core powered' model, heat from the core lifts the primordial atmosphere from within until it can escape the gravitational pull of the planet (Ginzburg et al. \citeyear{2018MNRAS.476..759G}; Gupta \& Schlichting \citeyear{2020MNRAS.493..792G}; Rogers et al. \citeyear{2021MNRAS.508.5886R}). Additionally, a range of other mechanisms have been proposed to explain the radius gap, including gas-poor formation (Lee et al. \citeyear{2014ApJ...797...95L}), rapid boil-off which is enhanced by stellar XUV once the stellar accretion disk disperses (Rogers et al. \citeyear{2024MNRAS.529.2716R}), and giant impacts, which have been shown to adequately strip large primordial envelopes (Inamdar \& Schlichting \citeyear{2015MNRAS.448.1751I}; Liu et al. \citeyear{2015ApJ...812..164L}). 

Photoevaporation makes clear and testable predictions about whether a planet should retain its envelope given its mass, initial envelope fraction, equilibrium temperature, and stellar irradiation history. Thus measurements of small, close-in transiting planet masses are potential tests of the photoevaporation model. A complication, however, is a high degree of uncertainty in a given planet's irradiation history. A planet's time-integrated XUV exposure or `fluence' is crucial in understanding how planetary atmospheres evolve in the photoevaporative model.

At a given stellar mass, the observed spread in pre-main-sequence rotation rates leads to a significant variance in stellar activity and, consequently, in the X-ray and ultraviolet (XUV) luminosity emitted by the star. This variability in $L_\mathrm{xuv}$ can result in a tenfold difference in the XUV exposure received by orbiting planets, depending on the stellar rotation rate and age  (Jackson et al. \citeyear{2012MNRAS.422.2024J}; Tu et al. \citeyear{2015A&A...577L...3T}). Owen \& Campos Estrada (\citeyear{2020MNRAS.491.5287O}) noted that multi-planet systems offer an opportunity to circumvent this uncertainty since all planets receive the same XUV fluence, up to a factor of $(a_{1}/a_{2})^2$ where $a_1$ and $a_2$ are the semi-major axes of the planets. They also noted that `disordered systems,' those with a sub-Neptune orbiting interior to a rocky super-Earth, provide the strongest constraints on the photoevaporative model.

In this study, we describe our efforts to measure the masses of the two transiting planets in the M-dwarf system TOI-1266. Early observations from the TESS Primary Mission (PM) supported a disordered configuration, suggesting a clear distinction between the two planets' sizes and compositions. However, subsequent analysis of data from the TESS Extended Mission (EM) has complicated this picture.

The transit depths of the planets, particularly the planet c, showed apparent variations across different observation sectors. These fluctuations have led to revisions in the estimated radii, with the once-reported super-Earth now more consistently aligning with the characteristics of a sub-Neptune (Cloutier et al. \citeyear{2024MNRAS.527.5464C}). Furthermore, Cloutier et al. (\citeyear{2024MNRAS.527.5464C}) reported evidence for a potential third candidate planet with a 32.3~d period in the system using HARPS-N radial velocity observations. 

Here, we present new RV observations for the TOI-1266 system using the M-dwarf Advanced Radial velocity Observer Of Neighboring eXoplanets (MAROON-X) spectrograph (\S \ref{sec:Observations}). We search for additional non-transiting planets but find none, and report refined mass measurements of the two transiting planets (\S \ref{sec:Results}). We interpret these planets in the context of other sub-Neptunes with well-constrained masses (\S \ref{sec:discuss}).

\begin{table*}[htb]
\centering
\caption{Stellar Parameters for TOI-1266}
\label{tab:table1}
\setlength{\tabcolsep}{10pt} 
\renewcommand{\arraystretch}{1.0} 
\begin{tabular*}{\textwidth}{@{\extracolsep{\fill}}cccc}
\hline
\text{Parameter} & \text{Unit} & \text{Value} & \text{Reference} \\
\hline
$T_{\text{eff}}$ & [K] & 3618 $\pm$ 157 & Cloutier et al. \citeyear{2024MNRAS.527.5464C} \\
Age & [Gyr] & 7.9 $\pm$ 5.2 & Stef\'ansson et al. \citeyear{2020AJ....160..259S} \\
Metallicity [Fe/H] & [dex] & --0.195 $\pm$ 0.115 & Cloutier et al. \citeyear{2024MNRAS.527.5464C} \\
$M_{*}$ & [$M_{\odot}$] & 0.431 $\pm$ 0.020 & Cloutier et al. \citeyear{2024MNRAS.527.5464C} \\
$R_{*}$ & [$R_{\odot}$] & 0.436 $\pm$ 0.013 & Cloutier et al. \citeyear{2024MNRAS.527.5464C} \\
$v \sin(i)$ & [km s$^{-1}$] & $< 1.3$ & Cloutier et al. \citeyear{2024MNRAS.527.5464C} \\
log $g$ & [log$_{10}$(cm s$^{-2}$)] & 4.73 $\pm$ 0.032 & Cloutier et al. \citeyear{2024MNRAS.527.5464C} \\
$\gamma$ & [km s$^{-1}$] & --41.639942 $\pm$ 0.00025 & Cloutier et al. \citeyear{2024MNRAS.527.5464C} \\
$\rho$ & [g cm$^{3}$] & 1.35 $\pm$ 0.36 & Cloutier et al. \citeyear{2024MNRAS.527.5464C} \\
$P_{\text{rot}}$ & [days] & 44.6 $\pm$ 0.8 & Cloutier et al. \citeyear{2024MNRAS.527.5464C} \\
Sp.T & & M3 V & Cloutier et al. \citeyear{2024MNRAS.527.5464C} \\
\hline
\end{tabular*}
\end{table*}

\begin{table}[!htb]
\centering
\caption{Summary of MAROON-X time series.}
\label{tab:table2}
\begin{tabular}{cccc}
\hline
\text{Date} & \text{$N_{obs}$} & \text{RMS} & $\gamma$\\
(UT) & (Red-Blue) & (m s$^{-1}$) & (m s$^{-1}$)\\
\hline
2022 March--April & 22 Red & 1.34 & $0.57 \pm 0.43$ \\
2022 March--April & 22 Blue & 1.79 & $1.89 \pm 0.48$\\
2022 May--June & 12 Red & 0.80 & $0.05 \pm 0.51$\\
2022 May--June & 12 Blue & 1.02 & $-1.21 \pm 0.54$\\
2022 July--July & 11 Red & 1.05 & $1.31 \pm 0.55$\\
2022 July--July & 11 Blue & 1.31 & $0.61 \pm 0.58$\\
2023 June--July & 18 Red & 1.16 & $-0.51 \pm 0.41$\\
2023 June--July & 18 Blue & 0.95 & $-0.80 \pm 0.44$\\
\hline
\end{tabular}
\end{table}

\section{Observations}
\label{sec:Observations}

We obtained 63 observations for TOI-1266 (Table \ref{tab:table1}) using the MAROON-X spectrograph, mounted on the 8.1m Gemini North telescope from 2022 March 24 UT to 2023 July 11 UT (program IDs: GN-2022A-Q-103, GN-2023A-Q-202). MAROON-X's thermal, mechanical, and optical stability, combined with its large collecting aperture, and red-optimized spectrometer make it an ideal instrument to study M-dwarf planetary systems. It uses both red (640--920~nm) and blue (490--720~nm) arms which yield two RV measurements per observation (Seifahrt et al. \citeyear{2018SPIE10702E..6DS}, \citeyear{2020SPIE11447E..1FS}, \citeyear{2022SPIE12184E..1GS}).

Our observational campaign spanned two observing seasons to maximize coverage and temporal resolution. Our sampling was tailored to the specific timing blocks during which MAROON-X was on sky. This resulted in 4 separate high cadence runs spanning from 2022A to 2023A. In total, we collected 126 RV measurements, with the data split evenly between MAROON-X's red and blue arms. The MAROON-X RV's are extracted using the SERVAL code which employs a template matching approach. A detailed overview of the template matching process and its application to precision radial velocity measurements is provided by Zechmeister et al. \citeyear{2020ascl.soft06011Z} and a full description of MAROON-X data reduction can be found in Winters et al. \citeyear{2022AJ....163..168W}

A summary of our time series analysis is detailed in Table \ref{tab:table2}. Across each epoch, we observed a consistent precision in our measurements, as evidenced by the RMS. For the red arm, the overall RMS aggregated across all intervals was 1.09~m~s$^{-1}$, with a median uncertainty of 0.89~m~s$^{-1}$. Similarly, for the blue arm, we achieved an overall RMS of 1.41~m~s$^{-1}$ and a median uncertainty of 1.14~m~s$^{-1}$.

\section{RV Analysis}
\label{sec:Results}
Cloutier et al. (\citeyear{2024MNRAS.527.5464C}) found evidence for a non-transiting third planet in the TOI-1266 system using the HARPS-N spectrograph. HARPS-N is an echelle spectrograph located at the 3.6m Telescopio Naxionale Gailieo (TNG) on La Palma, Canary Islands.  Here, we search the MAROON-X dataset for the presence of the third planet in the combined RV time series.

\subsection{Blind RV-Planet Search}


To do an independent search for the candidate planet reported by Cloutier et al. \citeyear{2024MNRAS.527.5464C} we computed a Generalized Lomb-Scargle periodogram on the MAROON-X dataset (Lomb \citeyear{1976Ap&SS..39..447L}; Scargle \citeyear{1982ApJ...263..835S}; Zechmeister and K\"urster \citeyear{2009A&A...496..577Z}). We applied eight independent offsets derived by the MAROON-X team by analyzing a group of well-behaved standard stars as described in Basant et al. (in prep). The MAROON-X offsets are as follows: 2022 March-April Red, Blue = $1.28\pm0.85$~m~s$^{-1}$, $2.53\pm1.10$~m~s$^{-1}$; 2022 May-June Red, Blue = $2.67\pm0.70$~m~s$^{-1}$, $1.29\pm0.58$~m~s$^{-1}$; 2022 July Red, Blue = $3.75\pm0.50$~m~s$^{-1}$, $2.43\pm.50$~m~s$^{-1}$; 2023 June-July Red, Blue = $4.08\pm0.56$~m~s$^{-1}$, $2.62\pm0.50$~m~s$^{-1}$.

\begin{figure}[!h]
 \centering
\includegraphics[width=\columnwidth]{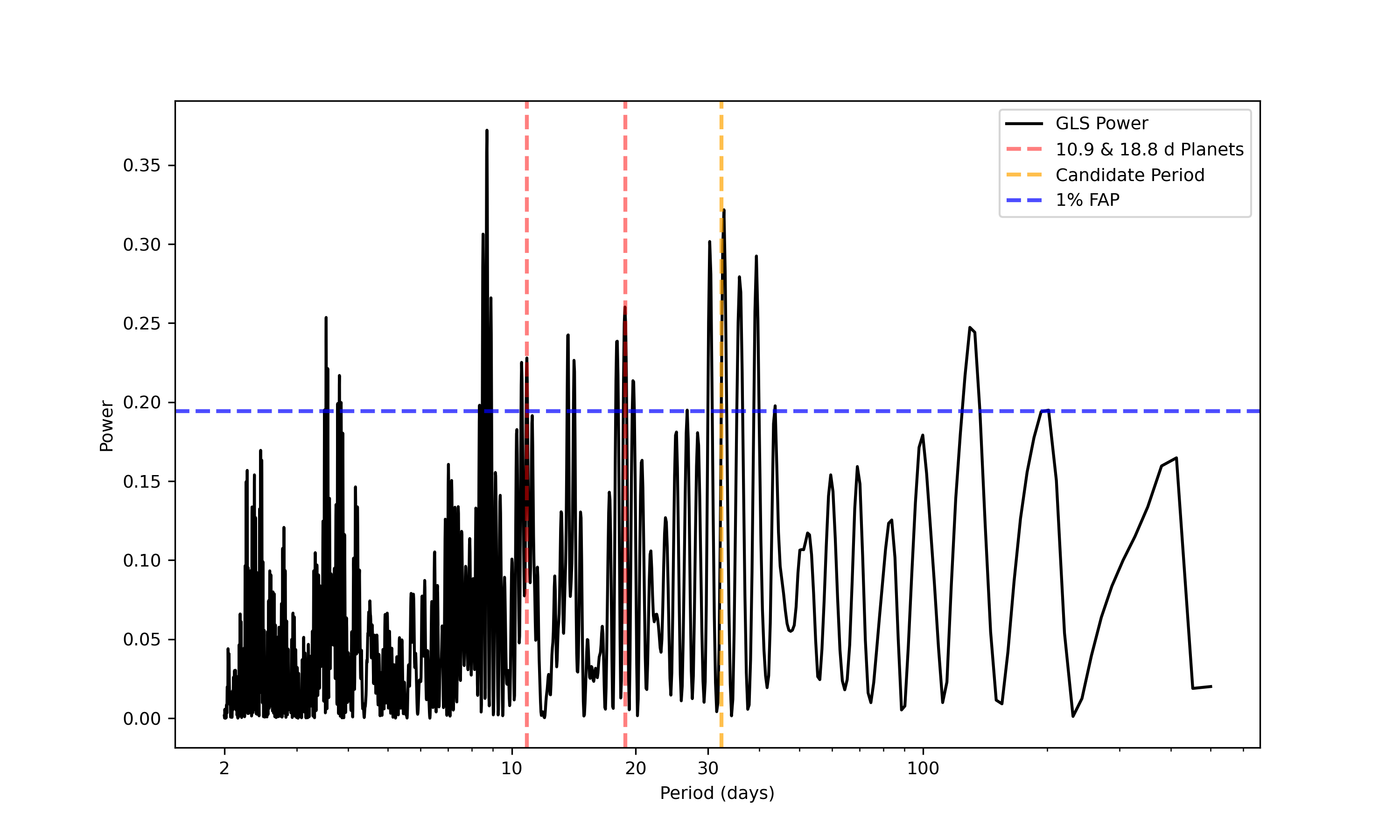}
\caption{Generalized Lomb-Scargle Periodogram for offset-corrected MAROON-X time series. The known planets periods are indicated with vertical red dashed lines ($P_{b}$ = 10.89 d) and 18.97 days ($P_{c}$ = 18.8 d). The candidate planet at 32.2~d is marked with a vertical orange dashed line.  The 1\% FAP is plotted with the horizontal blue dashed line.}
\label{fig:LS_Periodograms}
\end{figure}

In the blind RV search, we could not recover any planets which is not surprising due to our sampling and the number of independent offsets. We therefore conclude that we cannot confirm the candidate with our dataset. The periodogram can be seen in Figure \ref{fig:LS_Periodograms}.

\begin{figure*}[!htb]
 \centering
\includegraphics[width=\textwidth]{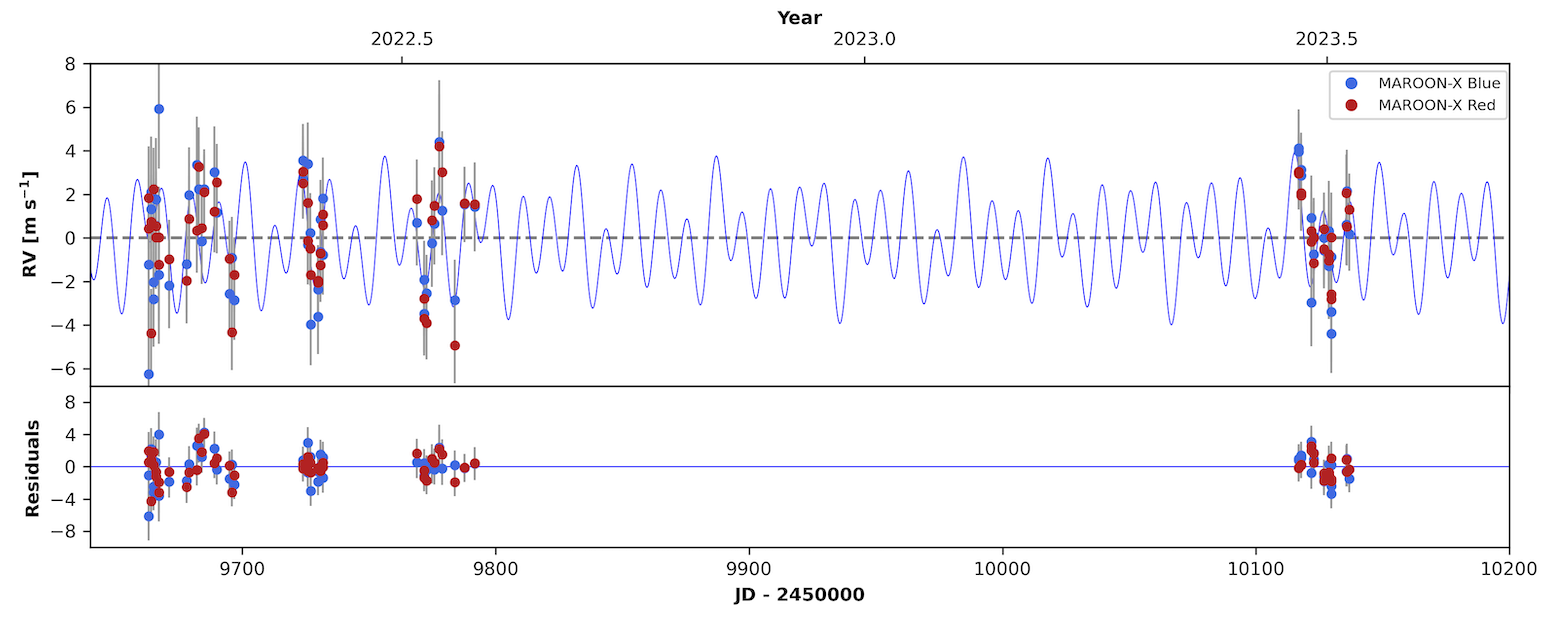}
	\caption{MAROON-X RVs collected from 2022 March to 2022 July during four separate observing runs, each with their own individual offsets. The blue points and red points represent the RVs from the blue and red arms, respectively. The best Keplerian model for the system is plotted with a solid blue line.}
	\label{fig:MaroonX_Rvs}
\end{figure*}

\subsection{Keplerian Orbit Fitting}
We measured the mass of planets b and c using \texttt{RadVel} (Fulton et al. \citeyear{2018PASP..130d4504F}), an open-source RV modeling pacakge. We adopted the orbital periods and conjunction times from Cloutier et al. (\citeyear{2024MNRAS.527.5464C}). We fixed the eccentricities to zero due to two main factors: 1) the RV semi-amplitudes were small relative to the per-measurement and error and 2) small planets in multi-planet systems exhibit small eccentricities with a median $e$ = 0.08 (Van Eylen et al. \citeyear{2019AJ....157...61V}; Yee et al. \citeyear{2021AJ....162...55Y}).

\begin{figure}[!h]
\includegraphics[width=1\columnwidth]{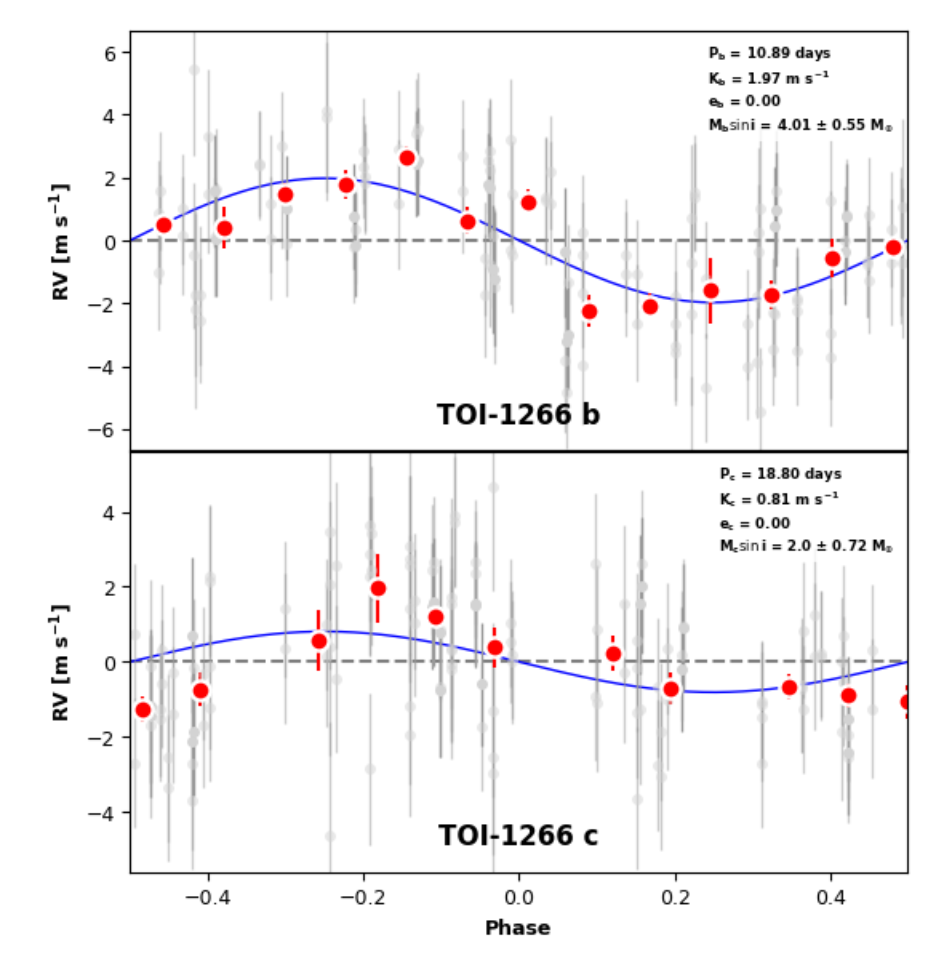}
\caption{Top panel: the RVs phase-folded at the period of TOI-1266 b, after subtracting the RV components from the other planet based on our best-fit model. The gray points represent the MAROON-X RVs. The model RV curve is plotted in blue. The phase-folded weighted mean RVs and their uncertainties are shown in red. Bottom panel: Same as top panel, but for TOI-1266 c.}
\label{fig:phasecurves}
\end{figure}

We included additional parameters in our Keplerian orbit analysis of the RVs: two RV jitter terms ($\sigma_{j}$) for both the red and blue arms which account for unmodeled white noise due to the star or instrument, and a 8 independent zero-point offsets ($\gamma$). MAROON-X is known to experience run-to-run offsets, which the MAROON-X team estimates with respect to the first observing run by observing a suite of standard stars. The $\gamma$ terms are different for the red and blue arms and for each of the four observing windows. However, to be conservative, we allow \texttt{RadVel} to treat each set of RVs as a different instrument and derive the best-fit off-set values which are listed in Table \ref{tab:table2}.

We performed a maximum likelihood estimation and then ran \texttt{RadVel}'s Markov Chain Monte Carlo (MCMC) algorithm (Foreman-Mackey et al. \citeyear{2013PASP..125..306F}) to explore the surrounding parameter space and estimate the uncertainties for our model parameters which can be seen in Table \ref{tab:table3}. Our full MAROON-X time series can be seen in Figure \ref{fig:MaroonX_Rvs}. 

Our Keplerian fit resulted in semi-amplitude values for the system of $K_{b} = 1.97 \pm 0.25$ m~s$^{-1}$ and $K_{c} = 0.81 \pm 0.29$ m s$^{-1}$ which correspond to masses of \Mbmx and \Mcmx. The phase-folded best fit models for each planet are shown in Figure \ref{fig:phasecurves}. 

We then combined our mass estimates with those of Cloutier et al. (\citeyear{2024MNRAS.527.5464C}) to refine the masses of these planets. For reference, the MAROON-X dataset included 126 total RV measurements with formal uncertainties of 0.89~m~s$^{-1}$ for the red arm and 1.14~m~s$^{-1}$ for the blue arm compared to 145 HARPS-N mass measurements with a median uncertainty of 1.91~m~s$^{-1}$. We report weighted average masses to be \Mbf and \Mcf. This results in density estimates of \rhob and \rhoc. For

\begin{table*}[htb]
\centering
\caption{Orbital Parameters and Planetary Properties Used, Measured, and Derived in This Work}
\label{tab:table3}
\setlength{\tabcolsep}{8pt} 
\renewcommand{\arraystretch}{0.9} 
\begin{tabular*}{\textwidth}{@{\extracolsep{\fill}}ccccc}
\hline
\text{Parameter} & \text{Unit} & \text{TOI-1266 b} & \text{TOI-1266 c} & \text{Reference} \\
\hline
$P$ & [days] & 10.894841 $\pm$ 0.000011 & 18.801611 $\pm$ 0.000053 & Cloutier et al. \citeyear{2024MNRAS.527.5464C} \\
$a$ & [au] & 0.0728 $\pm$ 0.0011 & 0.1047 $\pm$ 0.0016 & Cloutier et al. \citeyear{2024MNRAS.527.5464C} \\
$S_{p}$ & [$S_{\oplus}$] & 5.5 $\pm$ 1.1 & 2.7 $\pm$ 0.5 & Cloutier et al. \citeyear{2024MNRAS.527.5464C} \\
$T_{eq}$ & [K] & 425 $\pm$ 20  & 354 $\pm$ 16 & Cloutier et al. \citeyear{2024MNRAS.527.5464C} \\
$e$ & & 0 & 0 & This Work \\
$b$ & & 0.549 $\pm$ 0.057 & 0.769 $\pm$ 0.035 & Cloutier et al. \citeyear{2024MNRAS.527.5464C} \\
$i$ & [deg] & 89.13 $\pm$ 0.11 & 89.15 $\pm$ 0.06 & Cloutier et al. \citeyear{2024MNRAS.527.5464C} \\
$K$ & [m s$^{-1}$] & 2.16 $\pm$ 0.20 & 1.22 $\pm$ 0.22 & This Work \\
$M_{p}$ & [$M_{\oplus}$] & 4.39$^{+0.46}_{-0.45}$ & 2.97$^{+0.55}_{-0.55}$ & This Work \\
$R_{p}$ & [$R_{\oplus}$] & 2.62 $\pm$ 0.11 & 2.13 $\pm$ 0.12 & Cloutier et al. \citeyear{2024MNRAS.527.5464C} \\
$\rho$ & [g cm$^{3}$] & 1.35 $\pm$ 0.36 & 1.69 $\pm$ 0.40 & This Work \\
\hline
\end{tabular*}
\end{table*}

In combining the mass estimates, we report $>30\%$ improvements to the current mass uncertainties for the planets in the system.

\section{Discussion}
\label{sec:discuss}
\subsection{System Architecture}
Initially, the photometric observations for the system suggested planet c had radius $R_{c}=1.67~\pm~0.11~R_{\oplus}$ (Stef{\'a}nsson et al. \citeyear{2020AJ....160..259S}; Demory et al. \citeyear{2020A&A...642A..49D}). Both direct mass measurements of bulk planet densities (either by RVs or TTVs) and indirect evidence from population synthesis suggest that planets smaller than 1.7~$R_{\oplus}$ are preferentially rocky (see Lopez and Fortney \citeyear{2014ApJ...792....1L}; Rogers \citeyear{2015ApJ...801...41R}; Dressing and Charbonneau \citeyear{2015ApJ...807...45D}). In addition, population level analyses incorporating either photoevaporation or core-powered mass-loss suggest that planets below the radius gap are more likely to be rocky (Owen \& Wu \citeyear{2013ApJ...775..105O}, \citeyear{2017ApJ...847...29O}; Wu \citeyear{2019ApJ...874...91W}).

This seemingly positioned TOI-1266 as a rare disordered multi-planet system with a larger sub-Neptune orbiting interior to a a smaller rocky super-Earth. These disordered systems offer an opportunity to test the leading mass-loss theories by ensuring consistent radiation and formation environments (Owen \& Campos Estrada \citeyear{2020MNRAS.491.5287O}). If the outer planet is a stripped super-Earth, this sets a lower limit to the XUV fluence at its location of the inner sub-Neptune. By measuring the mass of the sub-Neptune, one may then test if the presence of a H/He envelope is consistent with the extrapolated fluence. 

As a low mass star TOI-1266 (0.4~$M_{\odot}$) was of particular interest because it has been demonstrated that the Radius Gap is less well-defined for planets orbiting M dwarfs (e.g. Petigura et al. \citeyear{2022AJ....163..179P}; Ho et al. \citeyear{2024MNRAS.531.3698H}). 

However, Cloutier et al. (\citeyear{2024MNRAS.527.5464C}) revised both planetary radii by carrying out an extensive analysis of additional TESS Extended Mission light curves and find that both planets are larger than originally reported. Cloutier et al. (\citeyear{2024MNRAS.527.5464C}) noted that the TESS transit depths from  TOI-1266 c's revised radius, $R_{c}=2.13\pm 0.12~R_{\oplus}$, suggests that the system, while still inverted, hosts two sub-Neptunes. Since the outer planet still retains its envelope, not much can be known about the XUV fluence at its location. Therefore, this system is not a strong test for photoevaporation or core-powered mass loss.

\begin{figure*}[!htb]
\centering
\includegraphics[width=1.0\textwidth]{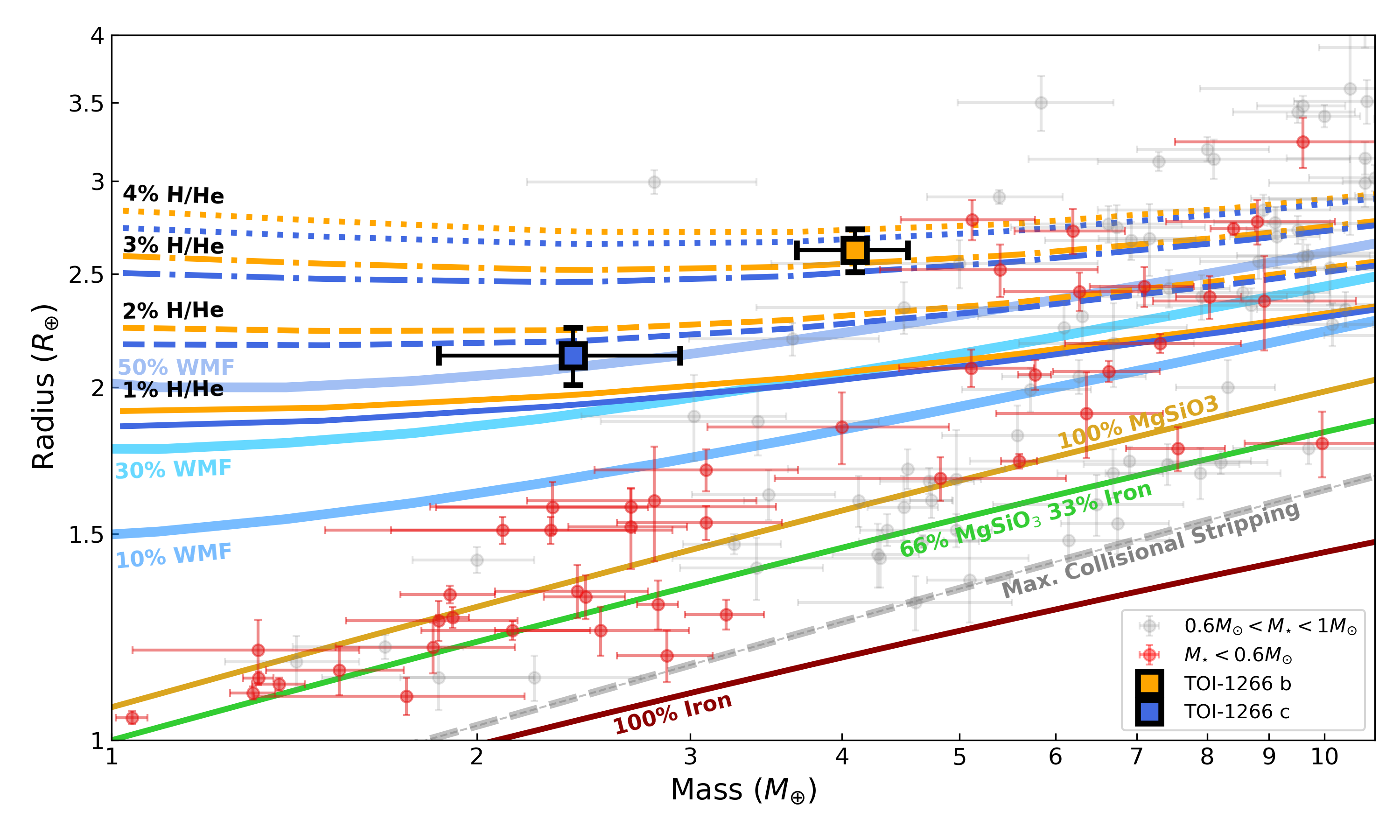}
\caption{Mass-Radius context figure showing the iso-composition curves and current known exoplanets up to 4~$R_{\oplus}$. TOI-1266 b and c and shown with gold and blue squares, respectively. The gray points represent composite data from the NASA Exoplanet Archive (\citeyear{nasa_exoplanet_archive}) filtered to only include planets that have published mass and radius uncertainties $<$ 30\%. The red data points represent such planets that orbit low mass stars ($M_{*} < 0.6~M_{\odot}$). The solid lines represent various iso-composition curves. The gold and blue 1\%--4\% H/He curves represent the predicted mass-radius curves for planets with various envelope mass fractions of \textit{exclusively} H/He for planets at $T_{eq,b}$~=~425~K and $T_{eq,c}$~=~354~K, respectively. The iso-composition curves for the solid structure models are from Zeng \& Sasselov (\citeyear{2013PASP..125..227Z}) and the envelope mass fraction curves were determined by interpolating in between pre-computed grids from thermal evolution models from Lopez \& Fortney (\citeyear{2014ApJ...792....1L}) that assume solar metallicity, H/He envelopes and ``Earth-like" cores, hot start thermal evolution and no atmospheric escape.}
\label{fig:isocurves}
\end{figure*} 

\begin{figure}[!htb]
\includegraphics[width=1.05\columnwidth]{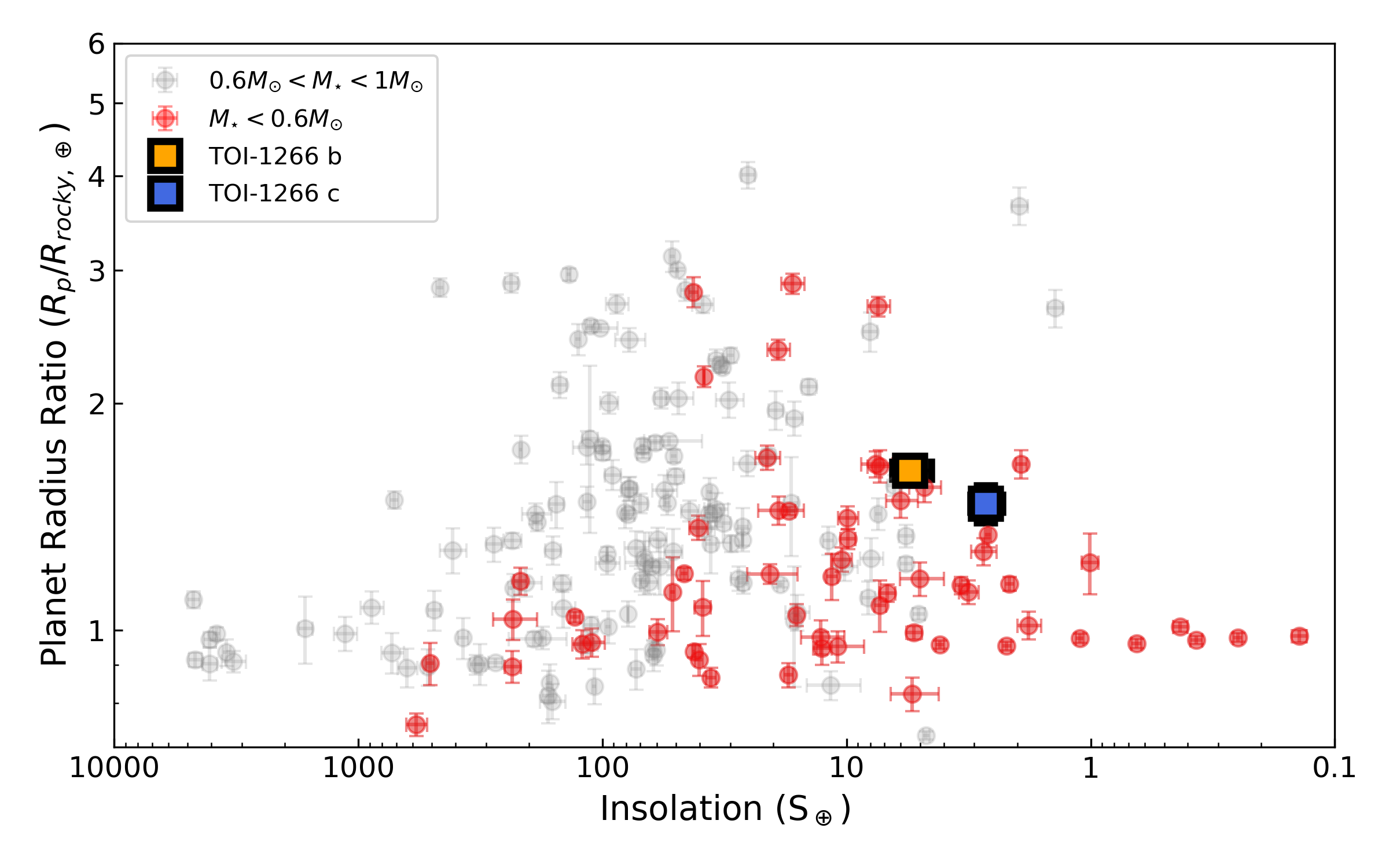}
\caption{Context plot showing the sample of exoplanets from the NEA in Figure~\ref{fig:isocurves}. Planets orbiting stars less than 0.6~$M_{\odot}$ are indicated in red. Along the x-axis we show insolation and on the y-axis we show the planet radius compared to the predicted radius for a rocky planet of the same mass.}
\label{fig:radiusratio}
\end{figure}

Cloutier et al. (\citeyear{2024MNRAS.527.5464C}) noted that the TESS Primary Mission (PM) transit depths from UT 2019 July 18 to 2020 March 17 (sectors 14, 15, 21, and 22) differed by 1.9$\sigma$ for planet b and 2.86$\sigma$ for planet c relative to transit depths measured in TESS Extended Mission (EM) from UT 2021 July 24 to 2022 March 25 (sectors 41, 48, and 49). Cloutier et al. (\citeyear{2024MNRAS.527.5464C}) performed extensive quality control checks to identify the source of this discrepancy and were able to rule out variable flux dilution, orbital precession, and residual artifacts in the light curve as possible sources. They did find some evidence that the star's magnetic activity increased during a later observing window but show that the impact would not likely explain the full change in transit depth. There is also some evidence that planet c's ephermeris would have been more accurately recovered in the EM relative to the PM, but note that this would also be insufficient in fully explaining the discrepancy. Although the culprit of the changing TESS depths could not be identified, the preponderance of data favors radii larger than 2~$R_{\oplus}$ for TOI-1266 c (Cloutier et al. \citeyear{2024MNRAS.527.5464C}).

\subsection{TOI-1266 b and c bulk compositions}
We plot the masses and radii of planet b and c in Figure~\ref{fig:isocurves} with the other known small planets with well-measured masses and radii compiled at the NASA Exoplanet Archive (NEA, \citeyear{nasa_exoplanet_archive}). We explored the bulk composition of these planets under the assumption that they are Earth-composition cores surrounded by envelopes of H/He. We consulted the Lopez \& Fortney \citeyear{2014ApJ...792....1L} models which tabulate how the radii of these two component planets vary with different H/He envelope fractions, age, equilibrium temperature, and mass. We interpolated these models adopting the $T_{\text{eq}}$ in Table \ref{tab:table3} and the age in Table \ref{tab:table1}. We found the envelope mass fractions for planet b and c to be 3.3$\pm$0.5\% and 1.6$\pm$0.4\% respectively. We note that these thermal evolution models assume no atmospheric escape. 

Both planets can be explained as having rocky cores with modest H/He envelopes. The isocomposition curves for 100\% MgSiO$_{3}$, 66\% MgSiO$_{3}$ and 33\% iron, 100\% iron, and maximum collisional stripping model predictions are from Zeng \& Sasselov \citeyear{2013PASP..125..227Z}. TOI-1266 b and c are both substantially lower density than one would predict for purely rocky planets and in the spread in radius at a given mass, both planets are located near the upper end of the distribution. Both planets require a low density component. In Figure \ref{fig:radiusratio} we show TOI-1266 b and c in a context plot with exoplanets with the same NEA filter as Figure~\ref{fig:isocurves} with the predicted radius ratio for rocky planets with the same measured masses.

\subsection{Theoretical Mass Loss Models}
\begin{figure}[h]
\includegraphics[width=1.0\columnwidth]{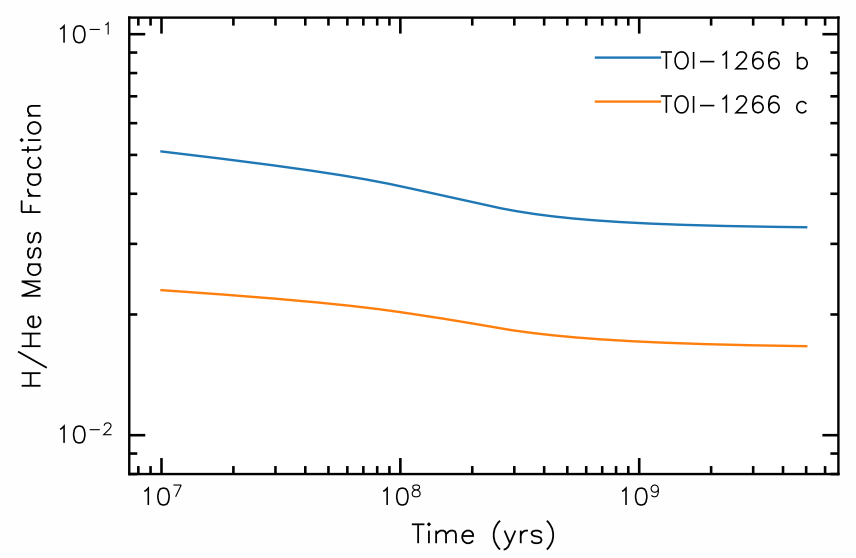}
\caption{Plausible evolution of envelope fraction over time given the current masses and radii of the TOI-1266 systems within the photoevaporative mass-loss framework of Rogers et al. (\citeyear{2021MNRAS.508.5886R}). Here planet b is more envelope rich than planet c at early and late times.}
\label{fig:mass_loss_theory}
\end{figure}

Based on the envelope mass fractions that we inferred using the models from Lopez \& Fortney (\citeyear{2014ApJ...792....1L}), we calculated theoretical mass loss tracks for each planet using the model from Rogers et al. (\citeyear{2021MNRAS.508.5886R}) which assumes nominal XUV stellar history tracks (Choi et al. \citeyear{2016ApJ...823..102C}; Dotter \citeyear{2016ApJS..222....8D}) and can be seen in Figure \ref{fig:mass_loss_theory}. Under these conditions, both planets would have experienced mild mass-loss. The higher current envelope fraction of the inner planet can be easily explained if the planet started out with a higher initial envelope fraction. 

\

\subsection{Water World Interpretation}
Our previous interpretations have assumed that the planets in this system are well-approximated by two volumetrically significant components: a mixed silicate-iron composition core and an envelope of H/He. However, one may replace some or all of the rock with water and explain the observed masses and radii. A number of groups have considered the possibility of water worlds. The high prevalence of low density super-Earths around M-dwarfs may be explained by irradiated ocean worlds with substantial water mass fractions (see: Cloutier \& Menou \citeyear{2020AJ....159..211C}; Luque \& Pall{\'e} \citeyear{2022Sci...377.1211L}; Cherubim et al. \citeyear{2023AJ....165..167C}; Piaulet et al. \citeyear{2023NatAs...7..206P}). In such systems, it is possible that planets that formed beyond the snow line and are comprised of volatiles later migrate inward and lose their gaseous envelopes boasting lower radii and densities.

The 10\%, 30\%, and 50\% theoretical water mass fraction lines for a planet of an equilibrium temperature of 400~K using the water-rich mass-radius relation models from Aguichine et al. (\citeyear{2021ApJ...914...84A}) can be seen in Figure~\ref{fig:isocurves}. Here, we can see that many super-Earths could also potentially be explained by rock+water. If indeed a rocky water-world, TOI-1266 c would be situated near the upper end of this distribution of planets with a water mass fraction near 50\%. This would likely imply separate formation pathways for both planets. While we cannot rule this out, we disfavor the water-world interpretation from an Occam's razor perspective compared to two planets that formed similarly and both retained modest H/He envelopes.

\section{Conclusion}
\label{sec:Conclusion}
In this study, we present 126 MAROON-X Doppler measurements of TOI-1266. We find masses consistent with those published by Cloutier et al. (\citeyear{2024MNRAS.527.5464C}). Combining these two independent measurements, we report refined masses of \Mbf~and \Mcf which require low density envelopes to explain their sizes. Indeed, the planets are among the most volatile-rich planets orbiting low-mass stars. Measurements of bulk density alone are insufficient to discriminate between two component models including 
rock \& H/He vs three-component models that include rock, water, and H/He. We favor the former interpretation because it is simpler. 

As new extreme precision Doppler facilities come online, more planets are amenable to precise mass measurements. We expect valuable insights will emerge from measurements of bulk density across diverse properties including planet size, system architecture (e.g., ordered vs disordered), and stellar properties. However, fundamental degeneracies will persist in the interpretation of bulk density when three or more components are viable. Independent observations such as transmission/eclipse spectroscopy may help break such degeneracies. 

\section*{Acknowledgments}
This research has made use of the NASA Exoplanet Archive, which is operated by the California Institute of Technology, under contract with the National Aeronautics and Space Administration under the Exoplanet Exploration Program. DT is supported in part by the Cota-Robles Fellowship at UCLA, which was instrumental in the advancement of this research. The University of Chicago group acknowledges funding for the MAROON-X project from the David and Lucile Packard Foundation, the Heising-Simons Foundation, the Gordon and Betty Moore Foundation, the Gemini Observatory, the NSF (award number 2108465), and NASA (grant number 80NSSC22K0117). The Gemini observations are associated with programs GN-2022A-Q-103 and GN-2023A-Q-202. Support for this work was provided by NASA through the NASA Hubble Fellowship grant number HF2-51559 awarded by the Space Telescope Science Institute, which is operated by the Association of Universities for Research in Astronomy, Inc., for NASA, under contract
NAS5-26555.

\newpage
\bibliography{Text}
\end{document}